\documentclass{article}
\usepackage{frascatiphys,epsf}
\begin{document}
\title{ 
THE THEORY OF CP-VIOLATION -- IN AS MUCH OF A NUTSHELL AS WILL FIT ON
8 PAGES
}
\author{
Patricia Ball\\
{\em IPPP, University of Durham, Durham DH1 3LE, UK} 
}
\maketitle
\baselineskip=11.6pt
\begin{abstract}
Do you know that CP violation is intrinsically linked to the scalar
sector of the Standard Model 
and its extensions? If yes, you need read no further
--- if no, you may turn over the titlepage and start reading now.
\end{abstract}
\baselineskip=14pt
\bigskip

It is difficult to do justice to a topic as vast and complex as 
CP-violation in a 30-minutes
conference talk --- and even more so in a 8-pages contribution to
the proceedings. Well, practitioners in teaching \& learning do know
that nothing is impossible, and so I shall try to stand up to the challenge and
concentrate on a less common
viewpoint on the subject than is to be found in most 
textbooks,\footnote{\footnotesize Everything you ever wanted to know about
  CP-violation (and more) can be found in Ref.\cite{books}.} in the
hope the reader may find it as entertaining as
enlightening.

It is actually very surprising that CP should be violated at all. Many
gauge-theories preserve C(harge conjugation symmetry) and P(arity)
naturally \& separately, the probably 
most prominent ones being (massless) QED and
  QCD. Even more contrived theories, especially
designed to violate parity, like the chiral gauge-theory
\begin{equation}\label{eq:0}
{\cal L} = -\frac{1}{4}\, F_{\mu\nu} F^{\mu\nu} + \bar\psi_L i \sigma
D \psi_L,
\end{equation}
where only the left-handed (Weyl) fermions $\psi_L$ interact with 
gauge-bosons,\footnote{\footnotesize Whereas their right-handed counterparts are
  ``sterile'' and hence omitted from the theory.}
 are still invariant under  CP transformations, which implies that
 {\it CP is a natural symmetry of massless gauge theories}.
So where does CP-violation come in?
The catch is that, as the mass term 
\begin{equation}
m\bar\psi \psi\equiv m (\bar\psi_L \psi_R + \bar\psi_R \psi_L)
\end{equation}
violates gauge-symmetry, it is forbidden in $\cal L$ and hence
left-handed fermions must be massless --- at obvious
variance with experiment. If the theory (\ref{eq:0}) is to serve as model for
parity-violating interactions, it has to be amended 
in some ingenious way as to give mass
to the fermions (and gauge-bosons), but at the same time preserve
gauge-invariance.

In the Standard Model (SM), this objective is being achieved by adding a
scalar (Higgs) sector which generates a nontrivial  
ground-state (vacuum) of the theory. In general, this vacuum-state is
less symmetric than the full theory --- a
phenomenon usually referred to as spontaneous symmetry breaking
(SSB), which in the case of gauge-theories is dubbed Higgs mechanism
and allows gauge-bosons (and chiral fermions) to become massive. 
The Lagrangian of the SM can be written as
\begin{equation}
{\cal L}_{\rm SM} = {\cal L}_{\rm gauge}(\psi_L,\psi_R,W,\phi) + {\cal L}_{\rm
  Higgs}(\phi) + {\cal L}_{\rm Yukawa}(\psi_L,\psi_R,\phi),
\end{equation}
where the first term on the right-hand side, the equivalent of
(\ref{eq:0}), contains the kinetic terms of
the fields involved, i.e.\ left- and right-handed fermions
$\psi_{L}$ and $\psi_R$, gauge-bosons $W$ and scalar (Higgs) fields
$\phi$, as well as their gauge-interactions. The second term is the
potential felt by the scalar fields and is responsible for some of
them to acquire a nonzero vacuum
expectation value (VEV) which gives rise to SSB.
The third term describes interactions between fermionic and
 scalar fields, which after SSB induce fermion mass terms. In the
SM, ${\cal L}_{\rm Higgs}$ is automatically 
CP-invariant,\footnote{\footnotesize The
  reason being that there is only one Higgs-doublet; CP-violation in
  ${\cal L}_{\rm Higgs}$ can
  occur, however, in models with more than one Higgs-doublet.} which
leaves us with ${\cal L}_{\rm Yukawa}$ as the only possible source of
CP-violation in the SM.\footnote{\footnotesize Note that the QCD $\theta$-term
  $\theta_{\rm QCD} g_s^2/(64\pi^2) 
\epsilon^{\mu\nu\rho\sigma}G^a_{\mu\nu}G^a_{\rho\sigma}$ can be set to
0 if all quarks are massless.}
It is given by
\begin{equation}\label{eq:1}
{\cal L}_{\rm Yukawa} = -\lambda_{ij}^d \bar Q^i_L\cdot \Phi d^j_R -
(\lambda_{ij}^d)^* \bar d^j_R \Phi^\dagger \cdot Q_L^i + \dots,
\end{equation}
where the indices $i,j$ run over the three generations and the 
dots denote terms with up-type quarks. $Q_L^i$ denotes the SU$_{\rm
  L}$(2) quark
doublet $(u_L^i,d_L^i)$ and $\Phi$ the SU$_{\rm L}$(2) Higgs doublet
$(\phi^+,\phi^0)$. The second term on the right-hand side of
(\ref{eq:1}) is the complex conjugate of the first one --- as required
by the condition that the Lagrangian be a Hermitian operator. 

So how does ${\cal L}_{\rm Yukawa}$ transform under CP?
The P-trans\-for\-ma\-tion exchanges L (left) and R (right) indices, the
C-transformation exchanges particles ($d$ etc.) and antiparticles
($\bar d$ etc.), so that
\begin{equation}
CP:~~\bar Q^i_L\cdot \Phi d^j_R \to \bar d^j_R \Phi^\dagger \cdot
Q_L^i.
\end{equation}
Comparing with (\ref{eq:1}), we see that ${\cal L}_{\rm Yukawa}$ is
CP-invariant if $\lambda\equiv\lambda^*$. Hence, a
necessary (but not sufficient) condition for CP-violation is that
the Yukawa couplings $\lambda^{u,d}$ are complex.

What does all that actually mean? Well, one conclusion is that 
{\it CP-violation happens in the scalar
  sector} --- at least in the SM. What about extensions?
  The statement stays evidently true for  ``simple''
  extensions of the SM with just an enlarged gauge- and
  scalar-field content (e.g.\ two Higgs-doublet model), and it also
  applies to theories where CP is not violated explicitly by complex
  couplings, but by spontaneous symmetry breaking --- which by
  definition is related to the scalar sector. What about
  supersymmetry? Again, CP is conserved in theories with unbroken
  SUSY, for the same reasons as above, but complex couplings occur
  after SUSY-breaking. Another conclusion is that studying
  CP-violation means probing the scalar sector --- which is also one
  of the main objectives of the Tevatron and the LHC. 
In this sense the measurement
  of CP-asymmetries in K and B decays is complementary to the direct
  searches for Higgs {\it et al.} at high-energy colliders.

What about CP-violation in the SM? Well, after SSB the Yukawa
couplings $\lambda^{u,d}$ induce
$3\times3$ mass matrices for $u$ and $d$-type quarks which are
eigenstates under weak interactions. If the theory is to be expressed
in terms of states with definite mass, these matrices have to be 
diagonalized. The resulting transformation from the basis of weak
eigenstates to that of mass eigenstates,
\begin{equation}
  u_i^{\rm (weak)} = U^{(u)}_{ij}u_j^{\rm (mass)},\qquad  
d_i^{\rm (weak)} = U^{(d)}_{ij}d_j^{\rm (mass)},
\end{equation}
has no effect on neutral interactions,\footnote{\footnotesize 
That is: there are no
  tree-level flavour-changing neutral interactions in the SM. Such
  interactions (e.g.\ $b\to s$) show up only at loop-level.} 
$\bar u_i^{\rm (weak)}
u_i^{\rm (weak)}\equiv \bar u_i^{\rm (mass)} u_i^{\rm (mass)}$, but
profoundly changes charged interactions:
\begin{equation}
\bar u_i^{\rm (weak)} d_i^{\rm (weak)}\to \bar u_i^{\rm (mass)}
(U^{(u)})^\dagger U^{(d)}d_i^{\rm (mass)}.
\end{equation}
The matrix $V\equiv (U^{(u)})^\dagger
U^{(d)}$ describes the strength of $d$-type
quarks decaying into $u$-type quarks and is nothing else but the well-known
CKM matrix. As $U^{(u,d)}$ just rotate the quark basis, they are unitary,
and so is $V$. Any $3\times3$ unitary matrix can be parametrised in
terms of three angles (the familiar Euler angles of
three-dimensional rotations) and six complex phases. In the present
case, however, not all six phases are physical: five of them can
be ``rotated away'' by redefining the phases of the quark fields ---
which leaves three angles and one phase to describe the CKM matrix
$V$. It is this complex phase that is the {\it one and only source of
  CP-violation in the SM}.

\begin{figure}[t]
$$\epsfxsize=0.5\textwidth\epsffile{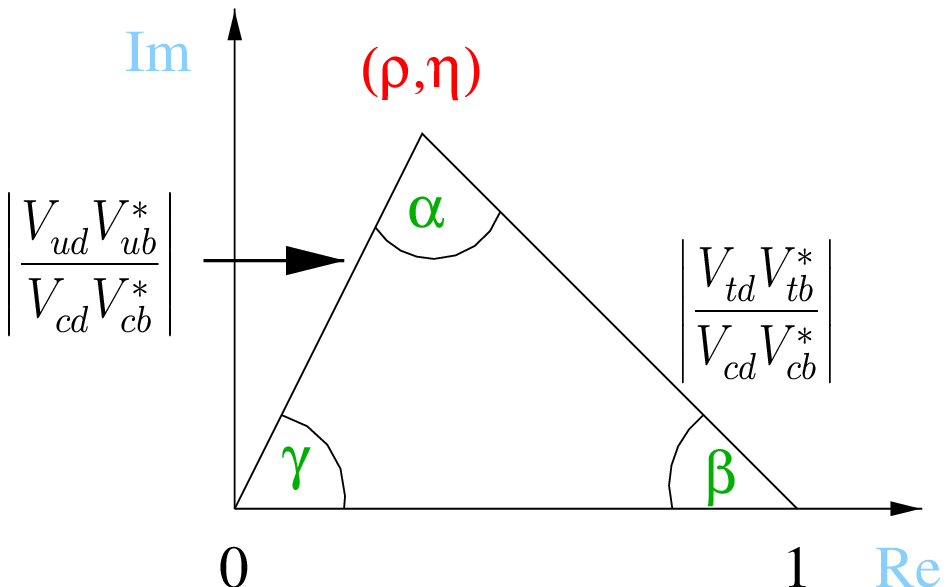}\quad
\epsfxsize=0.4\textwidth\epsffile{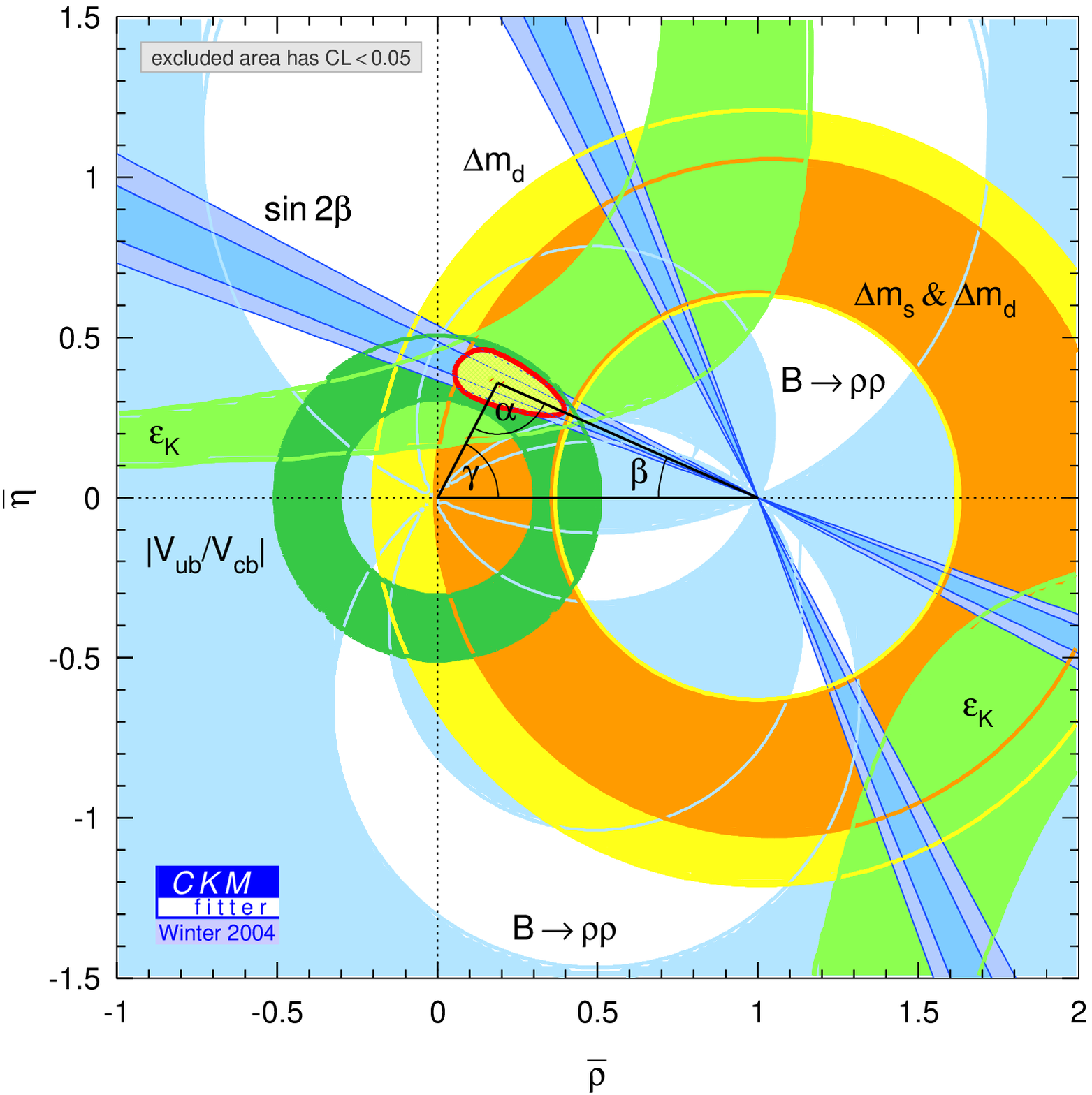}$$
\caption{\it Left: the $B_d$ unitarity triangle (UT). The apex is
  labelled $(\rho,\eta)$, which refers to the Wolfenstein
  parametrisation of the CKM-matrix. Right: the present
  (early 2004) experimental status of the UT\cite{ckmfitter}.}\label{fig:1}
\end{figure}

The fact that $V$ is unitary allows one to express the conditions for
CP-violation in the SM in an intuitively appealing form: unitarity
means
\begin{equation}
\sum_j V^{\vphantom{*}}_{ij} V^*_{kj} = \delta_{ik},
\end{equation}
which, for $i\neq k$, implies three complex numbers to add up to
zero. This relation can be represented by a triangle in the complex
plane, as shown in the left half of Fig.~\ref{fig:1}. For three
generations, there are six of these triangles in total. This
statement is true for arbitrary unitary matrices; the CKM matrix with
only one complex phase (instead of six in the general case) is
distinguished by the fact that all these six triangles have the same
area, which consequently is a 
measure of the strength of CP-violation in the SM.
{}From an experimental point of view, four of the triangles are
rather difficult to explore: one side is much smaller than the others,
which makes it difficult to measure the area (or angles) of these
triangles with sufficient precision. 
The two remaining triangles, with $i\in\{d,s\}$ and $k=b$, have
sides of comparable length, so that all their sides and angles are, in
principle, accessible in experiment: the 
$bd$ triangle is presently being studied at the B factories Babar and
Belle and its current experimental status is shown in
Fig.~\ref{fig:1}, 
right. The various constraints depicted in this figure are discussed in
other contributions to these proceedings. 
The $bs$ triangle will be the subject of experimental
scrutiny at the LHC. The objective of all these studies is to overconstrain the
triangles by measuring their sides and angles from various channels
and possibly refute the SM picture of CP-violation. Figure~\ref{fig:1}
shows that significant discrepancies yet have to be found.

The experimental determination of the sides and angles of the UT is
nothing less than trivial and will be the subject of 
other contributions to these
proceedings. Rather than embarking on a discussion of the respective merits
and shortcomings of various methods aiming to master the all-important
(and, in general,
yet unmastered) nonperturbative QCD effects in K and B 
decays, I would like to
spend the remaining three pages of this note on a discussion of the
bigger picture in which to embed any non-standard results on CP-violation.

So what are the alternatives to the SM picture of CP-violation? 
I  mentioned a few of them already; a more complete list includes
\begin{itemize}
\item complex couplings in the Higgs potential (e.g.\  
multi Higgs-doublet models);
\item complex couplings in the effective low-energy Lagrangian
  obtained from a fundamental theory by SSB (e.g.\ soft SUSY-breaking terms);
\item CP-violation from spontaneous symmetry breaking.
\end{itemize}
The latter scenario is rather attractive from the theorists' point of
view as it relieves us from the task of 
coming up with clever explanations
for where the complex couplings come from --- other than the standard
excuse that they are there because there is nothing to forbid them. 
If CP-violation is the result of SSB, the underlying
fundamental theory must be manifestly
CP-invariant, which requires the addition of (at least) an SU$_{\rm
  R}$(2) gauge-group to the SM. This type of theories goes by the name
of left-right symmetric models\cite{left-right} and has been studied
rather extensively. 
CP-violation occurs as consequence of
the SSB $SU_{\rm L}(2)\times SU_{\rm R}(2)\times U(1)\to SU_{\rm L}(2)\times
U(1)$. Like in the SM, fermion masses are generated from  
Yukawa interactions, but ${\cal L}_{\rm Yukawa}$ is now a bit more
involved and includes a Higgs-bidoublet $\Phi$, that is a doublet
under both SU$_{\rm L}$(2) and SU$_{\rm R}$(2).
CP-violation occurs as the VEV of $\Phi$ can carry a complex phase:
\begin{equation}
\langle\Phi\rangle = \left(\begin{array}{cc} v & 0\\ 0 & w
  \,e^{i\alpha}\end{array}\right).
\end{equation}
The phenomenology of this model has 
been recently studied in Ref.\cite{self}, for the quark sector; 
the main prediction,
a small value of $\sin 2\beta$, one of the angles of the $bd$ UT, has
not been confirmed by experiment. The other main prediction is large
CP-violation in
$B_s$ decays, which will be tested at the LHC. One major
problem of left-right symmetric models is the generically large
value of the electric dipole moment of the neutron, which is a two-loop
electroweak effect in the SM and hence exceedingly small, but occurs
at one-loop level and is dangerously large in left-right 
symmetric models
(and other models with additional sources of flavour-violation,
including SUSY). At present public opinion is rather in disfavour
of left-right models, but it is to be hoped that their more attractive
features, in particular the possibility of spontaneous CP-violation,
will eventually lead to their revival in an up-to-date form.

The last point I would like to stress in this note is the truly cosmic
implication of CP-violation: as Sakharov has
shown in 1967\cite{sakharov}, the fact that the Universe is dominated
by matter, and antimatter suspiciously absent, can only be explained if
\begin{enumerate}
\item fundamental interactions violate baryon number
  conservation;
\item the Universe has undergone non-equilibrium processes 
(phase-transitions) in its youth;
\item there is CP-violation, which allows Nature to distinguish
  baryons from antibaryons.
\end{enumerate}

\begin{figure}[t]
$$\epsfxsize=0.5\textwidth\epsffile{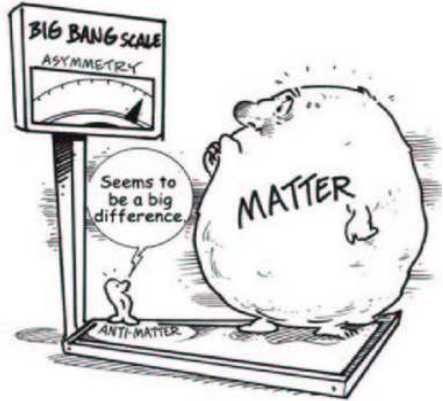}$$
 \caption{\it
      One rather weighty consequence of CP-violation.
    \label{fig:2} }
\end{figure}

Do we understand the origin of the cosmic matter-antimatter asymmetry?
Well, not {\it really}. Sacharov's conditions give us the minimum
ingredients, but don't tell us the recipe to use for cooking up the asymmetry.
Ever since 1967 creative {\it ma\^itres d'~} have come up with
ingenious compositions (e.g.\ GUT baryogenesis, leptogenesis, electroweak
baryogenesis), but none of them seems to get it quite right. One
result, however, does have emerged: 
CP-violation as observed in weak interactions is not
strong enough to explain the scale of the observed asymmetry --- which
leaves us with the exciting certainty that new physics must be out
there, longing to be discovered.

Let me conclude this {\it tour de force} 
by summarizing the messages I want to convey to you:
\begin{itemize}
\item CP-violation occurs in the scalar sector of the SM and its
  extensions;
\item in the SM, all CP-violation is related to one single complex
  phase in the CKM-matrix $V$;
\item the fact that $V$ is unitary and complex allows a simple
  visualisation of CP-violation in the SM: the unitarity
  triangle;
\item CP-violation is a phenomenon that does not only occur in the
  subatomic regime, but has profound consequences on the world we
  live in and is at the heart of the
  matter-antimatter asymmetry of the Universe. 
\end{itemize}

\section*{Acknowledgements}
I would like to thank the organisers of the workshop for providing a
pleasant and stimulating atmosphere. I also thank J.M.~Fr\`ere
who first introduced me to the marvels of CP-violation. 

\end{document}